\documentclass[useAMS]{mn2e}

\usepackage{graphicx}
\usepackage{epsfig}

\newcommand{\kms}{km~s$^{-1}$~}
\newcommand{\Teff}{T$_{\rm eff}$~}
\newcommand{\mkm}{$\mu$m~}
\newcommand{\vsini}{$v$~sin~$i$~}

\title[$V$~sin~$i$-s~ for late-type stars from spectral synthesis in K-band region.
]{
$V$~sin~$i$-s~ for late-type stars from spectral synthesis in K-band region.
}
\author[Yuri Lyubchik et al.]{
Yuri Lyubchik$^{1}$\thanks{E-mail:lyu@mao.kiev.ua}, 
Hugh R.A.Jones$^2$\thanks{E-mail:astqhraj@herts.ac.uk}, 
Yakiv V. Pavlenko$^{1,2}$\thanks{E-mail:yp@mao.kiev.ua}, 
David J. Pinfield$^2$\thanks{E-mail:D.J.Pinfield@herts.ac.uk}, 
\newauthor
Kevin R. Covey$^3$\thanks{E-mail:kcovey@astro.cornell.edu}\\
$^1$Main Astronomical Observatory of Academy of Sciences of
Ukraine, Zabolotnoho, 27, Kyiv, 03680 Ukraine\\
$^2$Center for Astrophysics Research, University of Hertfordshire,
College Lane, Hatfield, Hertfordshire AL10 9AB, UK\\
$^3$226 Space Sciences Building, Department of Astronomy,
Cornell University, Ithaca, NY 14853}

\begin{document}

\date{Accepted  . Received   ; in original form  }

\pagerange{\pageref{firstpage}--\pageref{lastpage}} \pubyear{ }

\maketitle

\label{firstpage}

\begin{abstract}

We analyse medium-resolution spectra (R$\sim$18000)
of 19 late type dwarfs in order to determine \vsini-s using synthetic  rather 
than observational template spectra. 
For this purpose observational data around 2.2 \mkm of stars with spectral classes 
from G8V to M9.5V were modelled.  

We find that the Na I ( 2.2062 and 2.2090~\mkm ) and~ $^{12}$CO 2-0 band features are 
modelled well enough to use for \vsini determination $without$ the need for a 
suitable observational template spectra. Within the limit of the resolution of our
spectra, we use synthetic spectra templates to derive \vsini values consistent with 
those derived in the optical regime using observed templates. We quantify the errors 
in our \vsini determination due to incorrect choice of model parameters
\Teff , log~$g$, $v_{\rm micro}$, [Fe/H] or FWHM and show that they are typically 
less than 10 per cent. We note that the spectral resolution of our data
($\sim$16~\kms) limited this study to relatively fast rotators and that resolutions 
of 60000 will required to access most late-type dwarfs. 
\end{abstract}


\section{Introduction}

The rotational velocity of a star is one of the basic parameters which can be obtained from
its spectrum. Much efforts has been devoted to the determination of surface
rotation of objects like protostars (Doppman et al., 2005) and main sequence stars  
with spectral types from mid-F (Charbonneau et al., 1997) through mid-M 
(Delfosse et al., 1998) until mid-L (Mohanty \& Basri, 2003) dwarfs.

Precise determination of stellar rotational velocities are useful for many different
analyses. Investigations have been devoted to the study of connection between \vsini and 
stellar activity (e.g., Noyes et al. 1984, Patten \& Simon 1996, Reiners \& Basri 2008). 
Stellar activity (chromospheric and coronal) increases with increasing of \vsini for
mid-F to mid-M dwarfs. However the rotational velocities distribution for stars later than M7V 
differs from the one for early types (e.g., Jenkins et al. 2009),
which is likely caused by the onset of fully convective interiors for spectral types
$\sim$ M3V and later (see e.g., Reiners \& Basri 2008). 
Moreover correlations of rotational velocity with 
spectral type and age were investigated 
both for stars of open clusters (Prosser et al., 1996; Stauffer et al., 
1994, 1997) and for field stars (Delfosse et al., 1998).

Most studies devoted to determination of rotational velocities used optical spectra 
and known 
velocity templates (Jenkins et al. 2009, West \& Basri 2009, Bailer-Jones 2004).
Recent progress in observational equipment provides spectra of 
relatively faint dwarfs in the infrared at high resolution and make it possible to use 
them in \vsini determinations (e.g., del Burgo et al. 2009, Reiners 2007, 
Lyubchik et al. 2007). 
Such studies should help provide a better description of stellar parameters
such as \Teff, log~$g$, atmospheric element abundances for late type dwarfs where they are 
rather uncertain (e.g. Jones et al. 2005,  Johnson \& Apps 2009). In this paper we focus 
on the derivation of rotational velocities in the infrared K band where there is a good 
match between observational and theoretical spectra. This match allows rotational velocities 
to be derived using theoretical spectra and to quantify the errors on derived stellar 
properties in a manner not feasible when using template velocities.

\section{Procedure}

\subsection{Observational spectra}

 For our derivation of rotational velocities, we used the spectra of late-type dwarfs 
presented in Doppmann et al. (2005).  
These near infrared spectra were obtained with the 10m Keck II telescope
on Mauna Kea, Hawaii using the NIRSPEC spectrograph during several observing runs. 
The details of the observations and reduction procedures 
are described in Doppmann et al. 2005. The resolution of the observed spectra is 
$\sim$18000 (Doppman et al. 2005), which corresponds to a rotational velocity of 
$\sim$16~\kms for the spectral regions of our interest. These  stars are used as 
standards in several studies of the properties of embedded young stars, 
including their evolutionary states (Doppmann et al. 2005), rotational and radial 
velocities (Covey et al. 2005, 2006), and magnetic field strengths 
(Johns-Krull et al. 2009).

\subsection{Synthetic spectra}

Synthetic spectra were computed using the WITA6 programme (Pavlenko, 2000). 
Calculations were carried out under the assumption of the local thermodynamic 
equilibrium, 
hydrostatic equilibrium and in the absence of sources and sinks of energy. 
The atomic line list used for our spectral modelling
and line identification is taken from the VALD (Kupka et al. 1999). 
The solar abundances reported by Anders \& Grevesse (1989) were used in calculations. 
All details of other input parameters are described by Pavlenko, Zapatero Osorio, 
Rebolo (2000). 

For the computation of the synthetic spectra we used molecular line lists from 
different sources:

$\bullet$Water vapour is the main contributor to the line opacity across most of the
infrared for late type M dwarfs and brown dwarfs.  There are several
H$_2$O line lists which are used at present time in computations of
synthetic spectra of dwarfs ( see for details and comparisons of H$_2$O line lists 
Pavlenko 2002, Jones et al. 2002, 2003, 2005 ). 
In our computations we used the BT2 line list (Barber et al. 2006) \\
$\bullet$ To model CO band  at $\sim$2.3~\mkm we used line list by 
Goorvitch (1994).\\

\begin{figure*}
\begin{center}
\includegraphics [width=160mm, height=140mm]{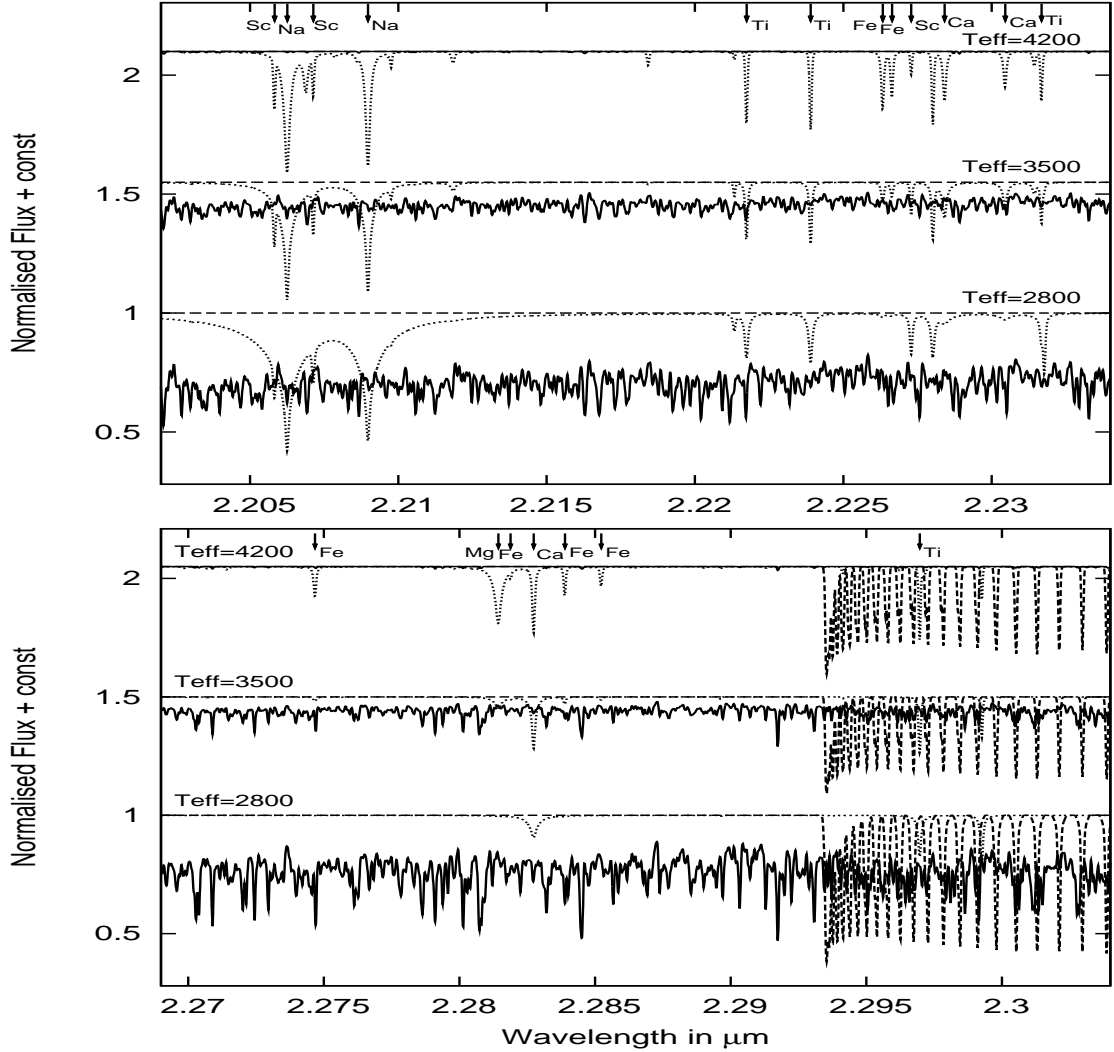}
\end{center}
\caption[]{\label{fig1}The relative contribution of different opacity sources 
to the stellar spectra is shown (atomic spectra are in dotted lines, molecular 
spectra of CO and H$_2$O in dashed and solid lines, respectively).}
\end{figure*}

The relative contributions of these molecules to the formation of
the spectra around 2.2~\mkm are shown in Fig. \ref{fig1}. This figure shows synthetic spectra 
with effective temperatures 2800~K, 3500~K, 4200~K, log~$g$=5.0, microturbulent
velocity $v_{\rm micro}$=2\kms, [Fe/H]=0.0.

The grid of synthetic spectra was computed from model structures of NextGen 
(Hauschildt  et al. 1999) for the following range of parameters: temperatures from 
2500 to 5000~K, log~$g$ from 4.5 to 5.5, microturbulent velocities from 1 to 3~\kms 
and metallicities from -0.5 to 0.0.

Theoretical spectra were computed with a wavelength step of 10$^{-5}$~\mkm and convolved
with Gaussians to match the instrumental broadening (R$\sim$18000). 
The rotational broadening of spectral lines is implemented following Gray (1976).

\subsection{Testing of model atmospheres parameters and synthetic spectra.}

In most papers the determinations of \vsini are made with reference to template 
spectra (e.g., Mohanty \& Basri 2003, Reiners \& Basri 2008). Such determinations 
are based on the assumption that the choice of rotational velocity standard is 
appropriate. However, in practice rotational velocity standards may have quite 
different physical properties though share similar spectra. 
 The primary goal of our paper is to obtain the rotational velocities using
synthetic spectra. For these purposes we 
investigate the affect of small changes in temperature, metallicity, 
gravity, microturbulent velocity and FWHM on \vsini determination in the spectral 
region of our interest 2.2670--2.3050~\mkm. 

Our tests are based on the automatic minimization procedure of Pavlenko \& Jones (2002).

Table~1 presents the results of analysis for synthetic spectra with effective 
temperatures 2800~K, 3500~K and 4200~K and \vsini-s 10, 20 and 35~\kms. Other 
model atmosphere parameters were fixed: log~$g$ = 5.0, $v_{\rm micro}$ = 2\kms and 
[Fe/H] = 0.0. Hereafter all these parameters are called reference ones.

First, we analysed how our determination of \vsini value reacts to increasing and 
decreasing of the FWHM parameter.
Synthetic spectra were computed with reference model atmosphere parameters and then 
convolved 
with fixed FWHM which correspond to the R = 18000 (16.65 \kms) in 2.2670--2.3050~\mkm 
wavelength region.
Then we changed the instrumental profile width within the limits of 10 per cent.
Deviations of the determined \vsini values from the reference \vsini-s are 
presented in column 3 of the Table~1.
Once the resolution limit is exceeded, the errors in rotational velocities 
determination depending on instrumental profile decrease to a few per cent.
Using synthetic spectra at \vsini = 10 \kms, the errors are higher, but remain less 
than 20 per cent, which is equal to 2 \kms, the average accuracy for \vsini 
determination with template spectra.

Other columns of the Table~1 present the percentage deviations of \vsini-s 
determination for model atmospheres $\Delta$-s of \Teff, 
log$g$,  $v_{\rm micro}$ and [Fe/H] from the reference \vsini-s.
Macroturbulent velocity is masked by using a Gaussian profile for the FWHM, so we didn't
analyse this parameter.
Errors in \vsini-s determination related to the deviations of model atmosphere 
parameters are larger than in the case of FWHM. Partially it is caused by
the relatively large step of parameters in available model atmospheres. 
Even so the errors are at the $\sim$10 per cent level with the only exception being the 
gravity parameter at 4200 K where the errors increase to 11-23 per cent.

Our investigations indicate that the errors for the reference value \vsini=10 \kms 
arise from minimization procedure errors while computing \vsini values less than 
instrumental resolution, so in the Table~1 the model differences are denoted 
in italic font.
Thus for the precise determination of rotational velocities we should use
the spectra with a resolution comparable or higher than the \vsini .

\begin{table*}
 \centering
 \begin{minipage}{140mm}
\caption[]{\vsini deviations based on an automatic minimization procedure for different model atmosphere parameters.}
 \label{del_param}
\begin{tabular}{c|c|c|c|c|c}
\hline
Reference model        & +/-0.1           & $\Delta$\Teff,   & $\Delta$log$g$,  & $v_{\rm micro}$      & $\Delta$[Fe/H]  \\
\Teff(K), \vsini(\kms) & of FWHM          &  +/-100K         &  +/-0.5dex       &  +/-1\kms        &  -0.5dex        \\                  
\hline
                       &    per cent      &      per cent    &    per cent      &  per cent        &  per cent \\
2800~~~~~~10           & {\it 18.0 / 10.0}& {\it 20.0 / 25.0}& {\it 15.0 / 5.0} & {\it 20.0 / 20.0}& {\it 10.0}   \\
~~~~~~~~~~~20          &  2.0  /   3.0    &  7.5 / 12.5      &  10.0 / 2.5      &  9.8 / 9.8       &  2.5       \\
~~~~~~~~~~~35          &  $<$1.0 /$<$1.0  &  5.7 / 8.6       &  5.7 / 2.9       &  4.2 / 7.0       &  2.9       \\ 
\hline
3500~~~~~~10           & {\it 14.6 / 12.5}& {\it 5.0 / 5.0}  & {\it 35.0 / 45.0}& {\it 20.0 / 25.0}& {\it 45.0}   \\
~~~~~~~~~~~20          &  2.0  /   2.0    &  2.5 / $<$1.0    &  10.0 /  20.0      &  10.0 / 10.0   &  15.0        \\
~~~~~~~~~~~35          &  $<$1.0 /$<$1.0  &  1.4 / 1.4       &  7.1 / 11.4        & 5.7 / 5.7      &  8.6       \\
\hline
4200~~~~~~10           & {\it 16.3 / 12.2}& {\it 5.0 /10.0}  & {\it 70.0 / 40.0}  &  {\it 20.0 / 20.0} &  (*)        \\
~~~~~~~~~~~20          &  2.0  /   2.0    &  2.5 / 5.0       &  22.5 / 17.5       &  10.0 / 10.0       &  (*)        \\
~~~~~~~~~~~35          &  $<$1.0 /$<$1.0  &  1.4 / 2.9       &  14.3 / 11.4       &  5.7 /  5.7        &  (*)        \\
\hline    
\end{tabular}
\end{minipage}\\
{ (*) -- we had no low metallicity model atmosphere for this temperature.}\\
\end{table*}

\section{Results of \vsini determinations.}

\begin{figure*}
\begin{center}
\includegraphics [width=160mm, height=150mm, angle=0]{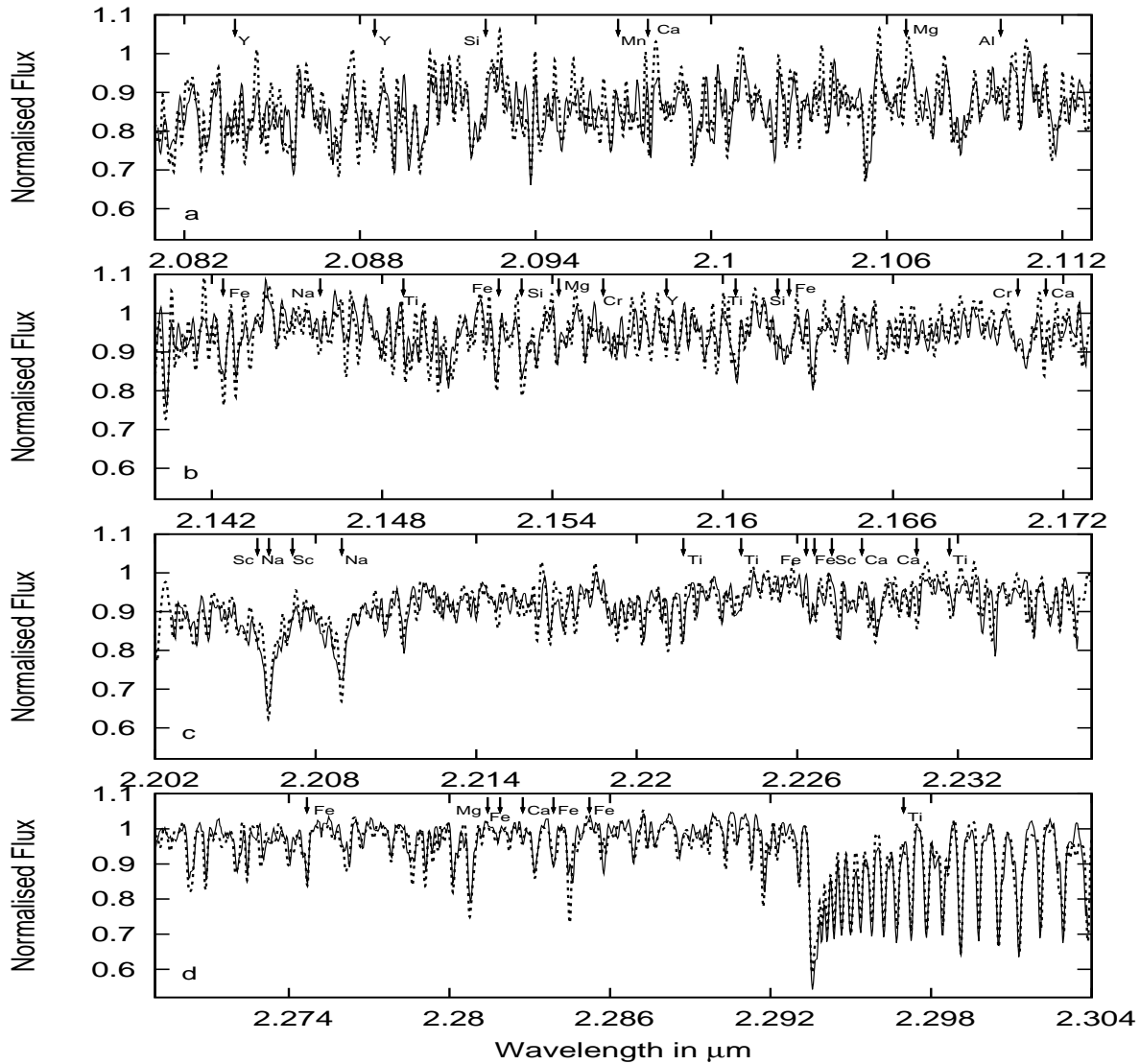}
\end{center}
\caption[]{\label{fig2}Synthetic spectra (dashed line) fitting to the observed spectra of VB~10 
(solid line) in the regions of interest.}
\end{figure*}

Using our grid of synthetic spectra we investigate the best fits to the observed 
spectra from Doppmann et al. (2005) which provide partial coverage 
from 2.0814 to 2.3040 \mkm ( 2.0800--2.1150~\mkm - Mg 2.1066~\mkm 
and Al 2.1099~\mkm strong lines;
2.1380--2.1750~\mkm - region containing of H~I Br$\gamma$ 2.1661~\mkm line;
2.2000--2.2390~\mkm - strong Na 2.2062 and 2.2090~\mkm lines; 
2.2670--2.3050~\mkm region with CO 2-0 band head at 2.2935~\mkm). 
We show the fits of synthetic spectra to the observed VB~10 (M5~V) in all 
spectral regions of interest as example on Fig.\ref{fig2}. 
The positions of atomic lines according to the VALD are labeled by arrows on 
the top of each panel.

\begin{figure*}
\begin{center}
\includegraphics [width=100mm, height=160mm, angle=270]{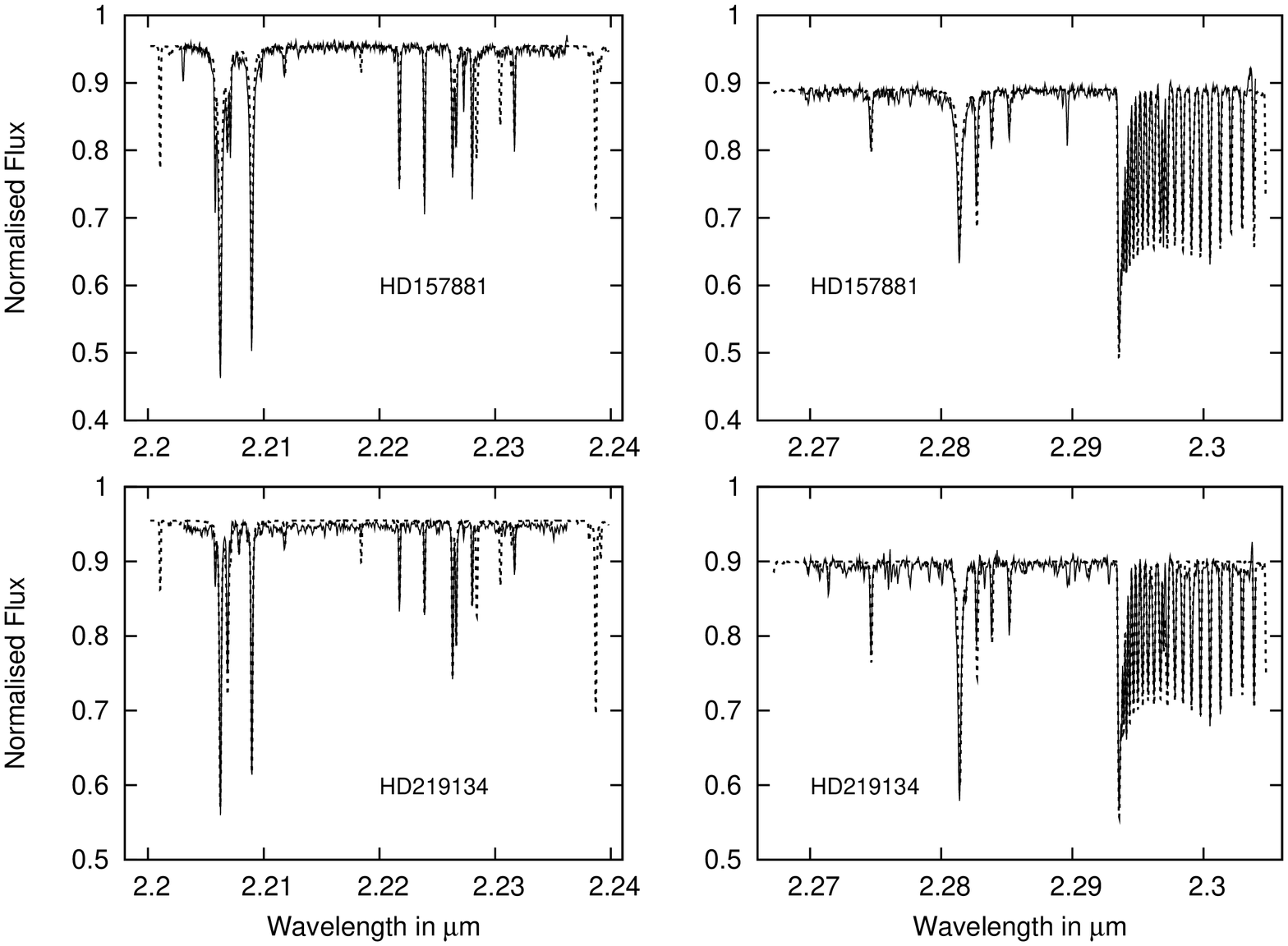}
\end{center}
\caption[]{\label{fig3}Synthetic spectra (dashed line) fitting to the observations 
(solid line) for the "Na" (left panels) and "CO" (right panels) regions for 
stellar spectra of K type stars .}
\end{figure*}

We found that our automatic procedure works better for hotter stars, where the 
contribution of strong molecular water lines is lower or has almost disappeared. 
However, 
for the cooler stars we found that the \vsini values derived from the Na lines
( 2.2062 and 2.2090~\mkm) and the CO band  
lines were usually considerably lower than those derived from other features. 
Rather than attribute such differences to astrophysical causes we believe this arises 
because the Na and CO features are intrinsically stronger 
than other spectral features and thus suffer less from blending with weak  worse modelled 
lines, mostly  H$_2$O.  We focus on two particular regions of interest:
2.2000--2.2390~\mkm (specifically around the strong Na doublet), hereafter "Na-region" and
2.2670--2.3050~\mkm (specifically around the region with CO band lines), "CO-region" 
hereafter in the text.  Figure \ref{fig1} shows these spectral ranges and the identification 
of other atomic lines in these regions: Fe, Sc, Ti, Mg, Ca and Cr. Fig. \ref{fig3} 
along with panels c and d of Fig.\ref{fig2} show 
the examples of best fit synthetic spectra for Na- and CO-regions for stars of different
spectral types.

Tables 2 and 3 present the effective temperature and \vsini determinations for 
the whole sample of stars from M9.5 to G8 spectral types for the different spectral 
regions.  The optical spectral and luminosity classes taken from Simbad database 
({\it http://simbad.u-strasbg.fr/simbad/}) are also shown.  In this work, initial 
effective temperature estimates were assigned on the based of spectral class using 
Tsuji et al. (1996) for M dwarfs, Allen (1973) for K dwarfs, and Doppmann \& Jaffe (2003) 
for G subdwarfs. Log~$g$ is defined by fitting to the wings of Na lines in observed 
spectrum.

 The best fit synthetic spectra are uncertain to within  
the range of $\pm\Delta$-s of model atmosphere parameters. The 
standard NextGen (Hauschildt  et al. 1999) model atmospheres grid steps 
$\Delta$\Teff = 100~K for the temperatures below 4000~K and 200~K above, 
$\Delta$log~$g$ = 0.5 and $\Delta$\vsini = 2\kms 
for determinations of upper limit values of \vsini are used. 
We used Solar metallicities for all stellar atmospheres calculations.

For stars with available from literature rotational velocity values, 
column (7) in the Tables 2 and 3 
gives the \vsini-s determined using templates. In Mohanty \& Basri (2003) 
(hereafter MB03) this was Gl406 with \vsini less than 3~\kms, in Marcy \& Chen (1992) 
(hereafter MCh92)  Gl~411 and Gl~820B with rotational velocities less than 2~\kms were 
used, so two \vsini values are given. In Duquennoy \& Mayor (1988) \vsini values were
obtained using the widths of the cross-correlation dips based on calibration of 
Benz \& Mayor (1981, 1984). The \vsini determination by using of cross-correlation 
function
was studied by other authors. Hartmann et al. (1986) and Rhode et al. (2001) 
indicate that 
this technique allows \vsini  determinations at the 90 per cent confidence level,
so the errors are less than 10 per cent.  

The differences in the derived model atmosphere parameters for some stars are at the same 
level as the model grid step. For \Teff and \vsini determinations of M dwarfs 
only strong atomic lines like Na~I 2.2062~\mkm, 2.2090~\mkm, and lines of 
CO molecule can be used. For hotter stars: Mg~I 2.2814~\mkm, 
Ti~I 2.2217, 2.2239, 2.2316, 2.2969~\mkm can be used.

\begin{figure*}
\begin{center}
\includegraphics [width=150mm, height=170mm, angle=270]{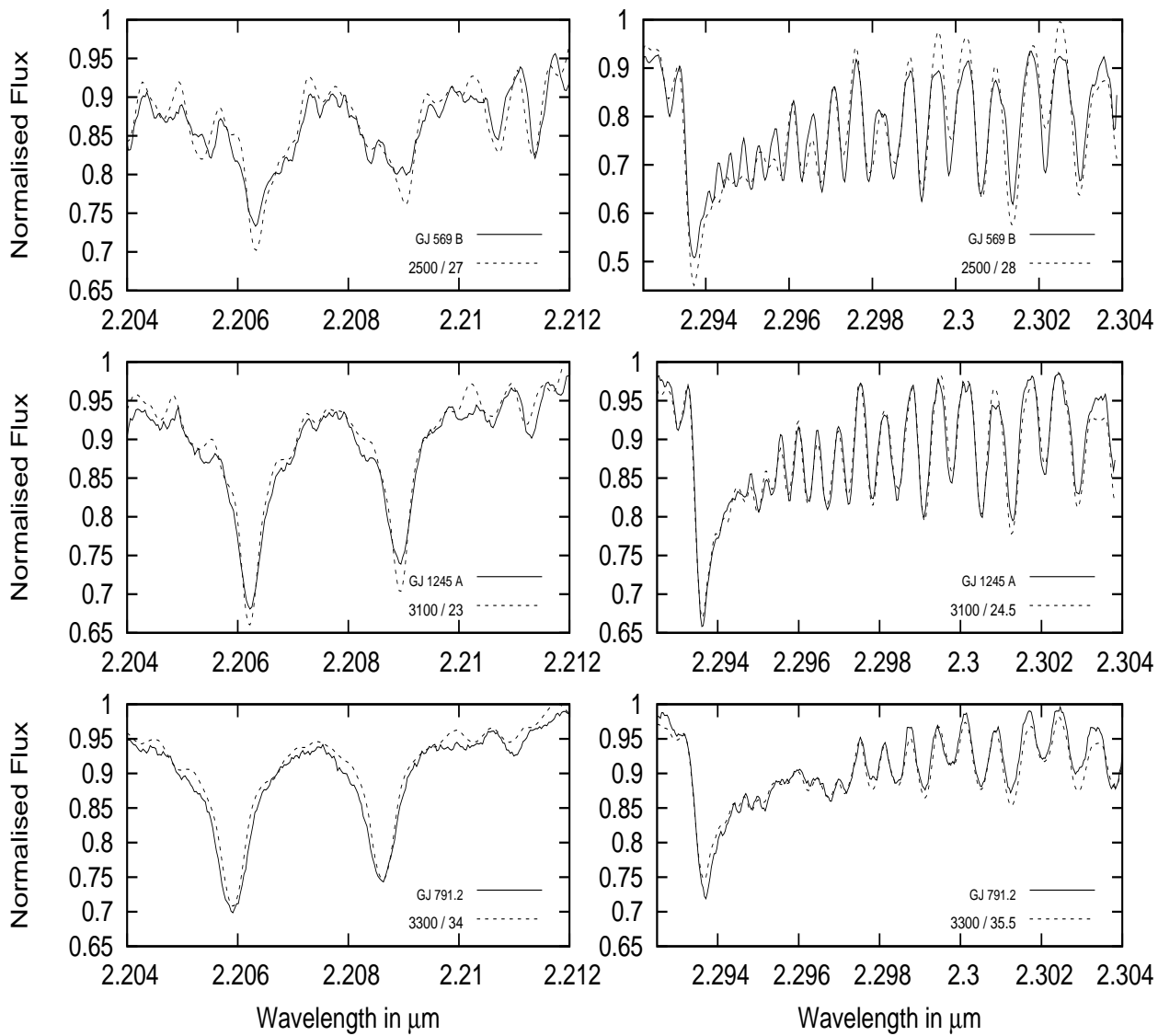}
\end{center}
\caption[]{\label{fig4}Synthetic spectra (dashed lines) of Na (left) and CO (right) lines fitting to the 
observed spectra (solid line). \Teff, K / \vsini, \kms are shown for modelled spectra.}
\end{figure*}

In Table 2 we present the \vsini determinations for three stars with rotational velocities 
greater than the instrumental resolution limit. For all stars we provide 2 values of \Teff 
and \vsini to show the impact of \Teff on \vsini determination. Features are shown in 
Fig.\ref{fig4}.
Looking at the sample as a whole in Table 2, it appears that Na and CO derivations of 
\vsini are consistent within the error bar with  values of Mohanty \& Basri (2003) 
derived using template spectra.
We can note a bit systematically higher value of \vsini-s derived from the "CO-region" than from 
Na lines. In the "CO-region" water lines are slightly stronger (see Fig.\ref{fig1}) and so
can add some 'extra broadening' to the CO features.  

 We should note that GJ569B and GJ1245A are known to be close orbiting 
binary systems and this may affect the observations analysed here. Our GJ569B 
spectra are formed by a near equal mass 
and spectral type components:  GJ569Ba M8.5V and GJ569Bb M9V (e.g., Lane et al. 2001). 
Our GJ1245A spectra consist of GJ1245A (M5.5V) and GJ1245C spectra. GJ1245C is 
3.3 magnitudes 
fainter in the V band (Henry et al. 1999) and so can only make a minor contribution 
at K band.

Table 3 shows the \vsini upper limit determination for 16 stars in a wide range of 
spectral 
classes from M9 to G8. We don't show the \vsini errors in columns 4 and 6, 
but from the analysis in section 2 we adopt that accuracy is about $\pm$2--4~\kms.
The usefulness of such upper limits values consists in \vsini determination through 
a wide range of spectral types using the synthetic spectrum method. 

\begin{table*}
 \centering
 \begin{minipage}{140mm}
\caption[]{\vsini determinations for the regions of the K band Na doublet and CO band.}
 \label{3stars}
\begin{tabular}{cc|cc|cc|cc}
\hline
         &         & "Na-region"   &              & "CO-region"    &              &  \vsini & Ref. \\ 
	 &         &  \Teff / logg & \vsini       & \Teff / logg   & \vsini       &  lit.   &         \\
\hline               
GJ569B   &  M8.5 V & 2400 / 5.0    & 19.0$\pm$6.5 & 2400 /5.0      & 32.0$\pm$6.5 &  ---    &      \\
         &         & 2500 / 5.0    & 27.0$\pm$1.0 & 2500 /5.0      & 28.0$\pm$1.0 &         &      \\
GJ1245A  &  M5.5 V & 3000 / 5.5    & 27.0$\pm$1.0 & 3000 / 5.5     & 27.0$\pm$1.0 &  22.5$\pm$2.0 & MB03 \\    
         &         & 3100 / 5.5    & 23.0$\pm$1.0 & 3100 / 5.5     & 24.5$\pm$1.0 &         &      \\    
GJ791.2  &  M4.5 V & 3200 / 5.5    & 31.5$\pm$3.0 & 3200 / 5.5     & 37.5$\pm$3.0 &  32$\pm$2.0 & MB03 \\
         &         & 3300 / 5.5    & 34.0$\pm$1.0 & 3300 / 5.5     & 35.5$\pm$1.0 &         &      \\
\hline
\end{tabular}
\end{minipage}\\
{MB03 - Mohanty \& Basri 2003}\\
\end{table*}

\begin{table*}
 \centering
 \begin{minipage}{140mm}
\caption[]{\vsini upper limit determinations for the regions of Na doublet and CO band. 
Spectral types for observed stars were taken from the Simbad database 
( http://simbad.u-strasbg.fr/simbad ).}
\label{16stars}
\begin{tabular}{cc|cc|cc|cc}
\hline
         &         & "Na-region"    &         & "CO-region"    &         &  \vsini        & Ref. \\ 
	 &         & \Teff / log$g$ & \vsini  & \Teff / log$g$ & \vsini  &  lit.          &       \\
\hline               
LHS2924  &  M9 V   &  2500/5.0      & {\it18} & 2600/5.0       & {\it14} &  11.0$\pm$2.0  & MB03 \\   
VB10     &  M8 V   &  2800/5.0      & {\it12} & 2800/5.0       & {\it10} &  6.5$\pm$2.0   & MB03 \\   
VB8      &  M6.5 V &  2900/5.0      & {\it10} & 2800/5.0       & {\it12} &  9.0$\pm$2.0   & MB03 \\   
GJ4281   &  M6.5 V &  2800/5.0      & {\it16} & 2900/5.0       & {\it13} &  7.0$\pm$2.0   & MB03 \\    
GJ569A   &  M2.5 V &  3500/5.0      & {\it12} & 3400/5.0       & {\it12} &  4$\pm$0.6/2.9$\pm$0.8 & MCh92 \\
GJ806    &  M1.5   &  3600/5.0      & {\it10} & 3600/5.0       & {\it12} &  1.5$\pm$0.8/3.7$\pm$1.6 & MCh92 \\    
HD131976 &  M1 V   &  3700/5.0      & {\it11} & 3700/5.0       & {\it13} &  10.6$\pm$2.5  & DM88 \\
HD184489 &  K5 V   &  4000/5.0      & {\it15} & 4000/5.0       & {\it15} &  & \\
HD201091 &  K5 V   &  4200/5.0      & {\it16} & 4200/5.0       & {\it15} &  & \\
HD157881 &  K5 IV  &  4200/5.0      & {\it10} & 4200/5.0       & {\it11} &  &\\
ADS14636 &  K5 V   &  4000/5.0      & {\it13} & 4200/5.0       & {\it14} &  & \\
HD219134 &  K3 V   &  4800/5.0      & {\it10} & 4800/5.0       & {\it10} &  & \\
HD168387 &  K2 III &  4400/4.0      & {\it14} & 4400/4.0       & {\it12} &  & \\
HD166620 &  K2 V   &  5000/5.0      & {\it12} & 5000/5.0       & {\it13} &  & \\
HD175225 &  G9 IV  &  4800/4.5      & {\it14} & 5000/5.0       & {\it13} &  &\\
HD182572 &  G8 IV  &  4800/4.5      & {\it13} & 5000/5.0       & {\it20} &  & \\
\hline
\end{tabular}
\end{minipage}\\
{DM88 - Duquennoy \& Mayor 1988; MCh92 - Marcy \& Chen 1992; MB03 - Mohanty \& Basri 2003}\\
\end{table*}

\section{Discussion}

Due to ongoing improvements in input data and 
synthetic spectra, modelling is becoming a precise method for the determination of stellar 
atmosphere parameters 
including rotational velocities.
Although there are many remaining uncertainties with abundance patterns ( Gustafsson 2004), 
3D modelling ( Pereira 2009), NLTE ( see Melendez et al. 2009), 
the synthetic spectra of solar type stars are relatively well determined. For the case of 
cool dwarf stars, the  conditions of fully convective 
atmospheres and dust formation means that still there are some uncertainties with the model 
structures (e.g., Pavlenko et al. 2007). 
The presence of many H$_2$O, FeH and other molecular lines in the infrared
region also makes spectral analysis and stellar parameter determination rather 
difficult (see e.g. del Burgo et al. 2009). For example, in the case of
cool dwarf atmosphere conditions broad wings of resonance lines are overlapped by blends
of numerous molecular lines. It is very challenging to analyse the pure shape of  
lines in such spectra and to separate the noise effects from the errors in molecular lines 
modelling. 

In this paper we show the perspective of K band region,
especially individual features like Na or CO lines, using for model 
atmosphere parameters determination. This region is suitable for such task
because only H$_2$O molecules play a significant role in formation of late type 
stars spectra here (see Fig.\ref{fig1} and Pavlenko et al. 2006). 
Around 2.2~\mkm there is a good match between observational and synthetic spectra. 
This circumstance makes the K band region preferable to the optical region where 
there is strong TiO and VO absorption and the J band (around 1.1~\mkm ),
where FeH and CrH along with H$_2$O make spectral analysis much more difficult.

Although there are no established standards for rotational velocity determinations
for late type stars, usually these values are determined using the spectrum
of a slowly rotating star  with a close spectral type as a template star.
Mohanty \& Basri (2003) used as template spectrum M6 dwarf Gl~406,
Marcy \& Chen (1992) spectra of two dwarfs: Gl~411 M2 and Gl~820B M0/K7. 
A choice of template can affect the result as it is seen in column 7 of Table 3.
On the other hand the synthetic spectra method depends mainly on input 
parameters. In section 2 we have shown that model atmosphere parameters deviations
in boundaries of standard model atmosphere grid steps
will bring errors of as much as 10 per cent in the region around 
2.2~\mkm.

For stars with rotational velocities above the 
resolution limit of observed data, our
determinations of \vsini and results using the template method are 
within the error bar (Table~2). 
Unfortunately the spectral resolution of our observations corresponds
to \vsini$\sim$16~\kms which allows us to perform the  accurate analysis only for 
the fast rotators. 
For determination of the rotational velocities of slow rotating M dwarfs 
with \vsini~$\sim$~5~\kms one should have
the observations with spectral resolution better than $\sim$60000.
It is worth noting that determination of rotational velocities on a level of a few 
\kms become complicated by uncertainties of model atmosphere parameters. Deviations 
of effective temperature, gravity, microturbulent or macroturbulent 
velocities and other parameters lead to errors of \vsini
determination at the level of 1--2~\kms. 
But in spite of these uncertainties we are sure that synthetic spectra method 
applied to suitable spectra in infra red region will be useful in the determination 
of \vsini (and other parameters). When the new generation of infrared
instruments, e.g. APOGEE-like projects (http://www.sdss3.org/surveys/apogee.php), 
start to work such techniques will enable reliable automated \vsini determination.

\section{Acknowledgements}

We would like to thank Greg Doppmann  
for acquiring and sharing the data, 
the authors of the PHOENIX model atmospheres for making their data available. 
We are very appreciate to the referee 
for his useful comments which improved this manuscript.
The authors thank the Royal Society and RoPACS network for support of travel 
and work.

\bsp

\label{lastpage}

\end{document}